\documentclass[11pt]{article}
\usepackage{amsmath}
\usepackage{amsfonts}
\usepackage{amssymb}
\begin{document}

\date{}
\title{\textbf{New Construction of $\Delta$-Operator in
Field-Antifield Formalism }}
\author{\textsc{P.M.~Lavrov}\thanks{E-mail: lavrov@tspu.edu.ru} and
\textsc{O.V.~Radchenko}\thanks{E-mail: radchenko@tspu.edu.ru}\\
\\\textit{Tomsk State Pedagogical University,}
\\\textit{634041 Tomsk, Russia}}
\maketitle

\begin{quotation}
It is proven that the nilpotent $\Delta$-operator in the
field-antifield formalism can be constructed in terms of an
antisymplectic structure only.
 \noindent \normalsize
\end{quotation}

\vspace{.5cm}

\section{Introduction}
The field-antifield formalism, developed by Batalin and Vilkovisky
\cite{bv}, provides a unique closed approach to covariant
quantization of general gauge theories, based on a special kind of
global supersymmetry, the so-called BRST symmetry \cite{brst}.
Generally speaking, the BRST symmetry is expressed in the nilpotency
property of odd second-order differential $\Delta$-operator
constructed explicitly \cite{bv} in the Darboux coordinates.
Investigation of geometrical meaning of the field-antifield
formalism \cite{geom} requires to consider all basic objects defined
invariantly on a supermanifold equipped with an antisymplectic
structure (for definition, see below). First of all it concerns the
$\Delta$-operator. Up to now two definitions of $\Delta$-operator
are known \cite{Kh,BB}. Khudaverdian's construction of
$\Delta$-operator includes an antisymplectic structure and a measure
density. The requirement of nilpotency  for $\Delta$-operator in
this approach  leads to a compatibility condition between these two
structures. Batalin-Bering's definition of $\Delta$-operator
involves additionally an odd scalar curvature. In this case the
compatibility condition between an antisymplectic structure and a
measure density can be omitted, but the definition operates with
three independent structures.

Our aim of this work is to study the re-definition of
$\Delta$-operator acting on an arbitrary antisymplectic
supermanifold.  The re-definition is based on the remarkable fact
proved in \cite{lr} that there exists the only symmetric connection
which is compatible with a given antisymplectic structure. Then
$\Delta$-operator can be constructed in the form likes to the usual
Laplacian in the Riemannian geometry.

The paper is organised as follows. In Sect.~2, we remind the
definition and basic properties of the antibracket using an
antisymplectic structure. In Sect.~3, we discuss the existence of
the unique symmetric connection compatible with a given
antisymplectic structure. In Sect.~4, we introduce the
$\Delta$-operator being nilpotent and written in terms of an
antisymplectic structure only.  In Sect.~5, we give a short summary.

We use the condensed notation suggested by DeWitt \cite{DeWitt} and
definitions and notations adopted in \cite{lr}. Derivatives with
respect to the coordinates $x^i$ are understood as acting from the
left and are standardly denoted by ${\partial A}/{\partial
x^i}$. Right derivatives with respect to $x^i$ are labeled by the
subscript $"r"$ or the notation $A_{,i}={\partial_r A}/{\partial
x^i}$ is used. The Grassmann parity of any quantity $A$ is denoted
by $\epsilon (A)$.
\\

\section{Antisymplectic structure}
In this section, we remind the notation of an antisymplectic
structure which is used for construction of the antibracket being
the basic object in the field-antifield quantization of general
gauge theories in Lagrangian formalism \cite{bv}.

Let us consider a supermanifold $M$ with an even dimension $dim M=
2N$. In the vicinity of each point $P\in M$  local coordinates $x^i,
\epsilon(x^i)=\epsilon_i$ can be introduced. Let $\Omega^{ij}$
($\epsilon(\Omega^{ij})=\epsilon_i+\epsilon_j+1$) be an odd
non-degenerate second-rank tensor field of type $(2,0)$ on $M$
obeying the generalized symmetry property
\begin{eqnarray}
\label{Omegasym} \Omega^{ij}=(-1)^{\epsilon_i\epsilon_j}\Omega^{ji}
\end{eqnarray}
and satisfying the relations
\begin{eqnarray}
\label{OmegaJI} \Omega^{in}\frac{\partial\Omega^{jk}}{\partial
x^n}(-1)^{\epsilon_i(\epsilon_k+1)} +cycle(i,j,k)\equiv 0.
\end{eqnarray}
The inverse tensor field $\Omega_{ij}$ has the generalized symmetry
property
\begin{eqnarray}
\label{Omegainvsym}
\Omega_{ij}=(-1)^{\epsilon_i\epsilon_j}\Omega_{ji}.
\end{eqnarray}
and the relations (\ref{OmegaJI}) can be rewritten in terms of
$\Omega_{ij}$ as (see, \cite{lr})
\begin{eqnarray}
\label{OmegainvJI} \Omega_{ij,k}(-1)^{\epsilon_k(\epsilon_i+1)}
+cycle(i,j,k)\equiv 0.
\end{eqnarray}
Any odd non-degenerate tensor field $\Omega^{ij}$ obeying the
properties (\ref{Omegasym}) and (\ref{OmegaJI}) (or, equivalently,
$\Omega_{ij}$ with the properties (\ref{Omegainvsym}) and
(\ref{OmegainvJI})) is referred as an antisymplectic structure on a
supermanifold $M$.

Having an antisymplectic structure $\Omega^{ij}$, one can define the
following odd bilinear operation for any two scalar functions $A$
and $B$ on $M$
\begin{eqnarray}
\label{AB} (A,B)=\frac{\partial_r A}{\partial x^i}(-1)^{\epsilon_i}
\Omega^{ij}\frac{\partial B}{\partial x^j},\quad \epsilon((A,B))
=\epsilon(A)+\epsilon(B)+1.
\end{eqnarray}
This definition leads to the invariance of the operation (\ref{AB})
under local coordinate transformations $x\rightarrow {\bar x},\;
({\bar A},{\bar B})=(A,B)$. From Eqs. (\ref{Omegasym}) and
(\ref{AB}) it follows
\begin{eqnarray}
\label{ABsym} (A,B)=-(-1)^{(\epsilon(A)+1)(\epsilon(B)+1)}(B,A),
\end{eqnarray}
i.e., the generalized symmetry property for the operation
introduced. In its turn, from Eq. (\ref{OmegainvJI}) one can obtain
the following identity
\begin{eqnarray}
\label{ABJIsym} (A,(B,C))(-1)^{(\epsilon( A)+1)(\epsilon(
C)+1)}+cycle(A,B,C,)\equiv 0\;,
\end{eqnarray}
i.e., the Jacobi identity for the operation. It means that $(A,B)$
(\ref{AB}) is the antibracket. The pair $(M, \Omega)$ defines an
antisymplectic supermanifold.

\section{Connection}
We now equip an antisymplectic supermanifold with a symmetric
connection (a covariant derivative). Let an antisymplectic structure
$\Omega^{ij}$ be covariant constant
\begin{equation}
\label{Omegacon} \Omega^{ij}\nabla_k=0.
\end{equation}
Then the inverse tensor field $\Omega_{ij}$ will be covariant constant too
\begin{equation}
\label{Omegainvcon} \Omega_{ij}\nabla_k=0\;, \quad
\Omega_{ij,k}-\Omega_{il}\Delta^l_{\;jk} -\Omega_{jl}\Delta^l_{\;ik}
(-1)^{\epsilon_i\epsilon_j}=0\;,
\end{equation}
where $\Delta^i_{\;jk}$ ($\epsilon(\Delta^i_{\;jk})=
\epsilon_i+\epsilon_j+\epsilon_k)$ is a symmetric connection and the
symmetry property of $\Omega_{ij}$ (\ref{Omegainvsym}) was used. In
\cite{lr} it was proven that  there exists the unique symmetric
connection $\Delta^i_{\;jk}$ compatible with a given antisymplectic
structure. The result is
\begin{eqnarray}\label{OmegaDelta}
\Delta^l_{\;ki}=\frac{1}{2}\Omega^{lj}\Big(\Omega_{ij,k}
(-1)^{\epsilon_k\epsilon_i}
+\Omega_{jk,i}(-1)^{\epsilon_i\epsilon_j}-
\Omega_{ki,j}(-1)^{\epsilon_k\epsilon_j}\Big)(-1)^
{\epsilon_j\epsilon_i+\epsilon_l}.
\end{eqnarray}
We see that in the case of an antisymplectic supermanifold
$(M,\Omega)$ there is the only symmetric connection compatible with
a given antisymplectic structure. It looks like Riemannian
geometry.

Let us consider the antisymplectic curvature tensor
\begin{eqnarray}
\label{Rs}
{\cal R}_{ijkl}=\Omega_{in}{\cal R}^n_{\;jkl},\quad
\epsilon({\cal R}_{ijkl})=\epsilon_i+
\epsilon_j+\epsilon_k+\epsilon_l+1,
\end{eqnarray}
where ${\cal R}^n_{\;\;jkl}$ is the curvature tensor of a symmetric
connection $\Delta^i_{\;jk}$. This leads to the following
representation,
\begin{eqnarray}\label{r1}\nonumber
{\cal R}_{nljk}&=&-\Delta_{nlj,k}+\Delta_{nlk,j}(-1)^{\epsilon_j\epsilon_k}
+\Delta_{ink}\Delta^i_{\;lj}(-1)^
{\epsilon_n\epsilon_i+\epsilon_k(\epsilon_i+\epsilon_l+\epsilon_j)}\\
&&-\Delta_{inj}\Delta^i_{\;lk}(-1)^{\epsilon_n\epsilon_i+\epsilon_j
(\epsilon_i+\epsilon_l)}\;,
\end{eqnarray}
where
\begin{eqnarray}\label{D1}
\Delta_{ijk} =\Omega_{in}\Delta^n_{\;jk}\;,\quad\epsilon(\Delta_{ijk})
=\epsilon_i+\epsilon_j+\epsilon_k +1.
\end{eqnarray}
Using this, the relation (\ref{Omegainvcon}) reads
\begin{eqnarray}
\label{Omegainvcon1} \Omega_{ij,k}=\Delta_{ijk}+
\Delta_{jik}(-1)^{\epsilon_i\epsilon_j}.
\end{eqnarray}
This relation plays very important role in deriving the symmetry
properties of the antisymplectic curvature tensor \cite{lr}.

The antisymplectic curvature tensor obeys the following generalized
symmetry properties
\begin{eqnarray}
\label{Rans1} {\cal R}_{ijkl}=-(-1)^{\epsilon_k\epsilon_l}{\cal
R}_{ijlk},\quad {\cal R}_{ijkl}=-(-1)^{\epsilon_i\epsilon_j}{\cal
R}_{jikl}\;, \quad {\cal R}_{ijkl}={\cal
R}_{klij}(-1)^{(\epsilon_i+\epsilon_j) (\epsilon_k+\epsilon_l)}.
\end{eqnarray}
and the Jacobi identity
\begin{eqnarray}
\label{Rjac1} (-1)^{\epsilon_j\epsilon_l}{\cal R}_{ijkl}
+(-1)^{\epsilon_l\epsilon_k}{\cal R}_{iljk}
+(-1)^{\epsilon_k\epsilon_j}{\cal R}_{iklj}=0\,.
\end{eqnarray}

Ricchi tensor can be defined by contracting two indices of curvature tensor
\begin{eqnarray}
\label{Ricdef} {\cal R} _{ij}= {\cal
R}^k_{\;\;ikj}\;(-1)^{\epsilon_k(\epsilon_i+1)}= \Omega^{kn}{\cal
R}_{nikj} (-1)^{\epsilon_i\epsilon_k+\epsilon_k+\epsilon_n}, \quad
\epsilon({\cal R}_{ij})=\epsilon_i+\epsilon_j\;.
\end{eqnarray}
Ricchi tensor is a generalized symmetric tensor
\begin{eqnarray}
\label{Ricsym}
{\cal R}_{ij}={\cal R}_{ji}(-1)^{\epsilon_i\epsilon_j}.
\end{eqnarray}
The further contraction between a given antisymplectic structure and
Ricci tensor gives scalar curvature
\begin{eqnarray}
\label{Scalcur}
{\cal R} = \Omega^{ji}{\cal R}_{ij}\;(-1)^{\epsilon_i+\epsilon_j},\quad
\epsilon({\cal R})=1
\end{eqnarray}
which, in general, is not equal to zero. Notice that the scalar
curvature tensor squared is identically equal to zero, ${\cal
R}^2=0$. The scalar curvature tensor was quite recently used in
\cite{BB} for generalization of Khudaverdian's construction of
$\Delta$-operator \cite{Kh}.

\section{$\Delta$-operator}

Having an antisymplectic structure  $\Omega^{ij}$ and the unique
symmetric connection (the covariant derivative) compatible with this
structure there is the intrinsic definition of an odd second-order
differential operator $\bigtriangleup$ (the odd Laplacian)
\begin{eqnarray}
\label{Delt} \bigtriangleup=\frac{1}{2}\Omega^{ij}\;\nabla_j\nabla_i
(-1)^{\epsilon_i+\epsilon_j}=\frac{1}{2}
\nabla_j\nabla_i\;\Omega^{ij}(-1)^{\epsilon_i+\epsilon_j}\;,\quad
\epsilon(\bigtriangleup)=1
\end{eqnarray}
acting (from the right) on any tensor field defined on a
supermanifold $M$ as a scalar operator. Note that the definition
(\ref{Delt}) is completely given in terms of an antisymplectic
structure and its partial derivatives only. The operator
$\bigtriangleup$ is obviously nilpotent
\begin{eqnarray}
\label{Deltnil} \bigtriangleup^2=0.
\end{eqnarray}
Action of $\bigtriangleup$ on product of two scalar function reproduces the antibracket
\begin{eqnarray}
\label{Deltant}(A\cdot B) \bigtriangleup =A\cdot (B \bigtriangleup)+
(A\bigtriangleup)\cdot B(-1)^{\epsilon(B)} +
(A,B)(-1)^{\epsilon(B)}.
\end{eqnarray}
Above mentioned properties of $\bigtriangleup$ allows us to speak of
new definition of the $\bigtriangleup$-operator in the
field-antifield formalism. The quantum master equation
\begin{eqnarray}
\label{Mast} \exp{\Big\{\frac{i}{\hbar}W\Big\}}\bigtriangleup=0
\end{eqnarray}
is not modified
\begin{eqnarray}
\label{Mast1} \frac{1}{2}(W,W)=i\hbar W\bigtriangleup
\end{eqnarray}
in contrast with \cite{BB}, where generalization of
$\bigtriangleup$-operator due to the odd scalar curvature led to
modification of the quantum master equation written in terms of the
antibracket.

\section{Conclusion}

We have introduced the definition of $\Delta$-operator in the
field-antifield formalism in terms of an antisymplectic structure
and its partial derivatives only. It became possible because of the
remarkable fact: there is the unique symmetric connection compatible
with a given antisymplectic structure. It was shown that the
$\Delta$-operator is nilpotent, reproduces the antibracket in
standard way, gives the usual form of the quantum master equation.

\section*{Acknowledgements}
The work was partially supported by grant for LRSS, project No.\
4489.2006.2. The work of PML was also supported by the INTAS grant,
project INTAS-03-51-6346, the RFBR grant, project No.\ 06-02-16346,
the DFG grant, project No.\ 436 RUS 113/669/0-3 and joint RFBR-DFG
grant, project No.\ 06-02-04012.

\end{document}